\documentclass[amssymb,aps, prl,superscriptaddress, 10pt]{revtex4-1}

\usepackage{amsmath}
\usepackage{fullpage}
\usepackage{amssymb}
\usepackage{amsfonts}
\usepackage{color}
\usepackage[pdftex]{graphicx}
\usepackage{epstopdf}
\usepackage{soul}
\usepackage{xcolor}
\usepackage{multirow} 
\usepackage{caption}
\usepackage{subcaption}

\newcommand{\Cjm}{C_{jm_z}}

\newcommand{\Ei}{\mathbf{E^{i}}}
\newcommand{\Es}{\mathbf{E^s}}

\newcommand{\Emp}{\mathbf{E}_{p,m_z}^{\lambda}}

\newcommand{\aje}{\alpha_{j,m_z}^{(e)}}
\newcommand{\ajm}{\alpha_{j,m_z}^{(m)}}

\newcommand{\Am}{\mathbf{A}_{jm_z}^{(m)}}
\newcommand{\Ae}{\mathbf{A}_{jm_z}^{(e)}}

\begin{document}

\title{Tailoring multipolar Mie scattering with helicity and angular momentum} 
\author{Xavier Zambrana-Puyalto}
\email{xavislow@protonmail.com}
\affiliation{Department of Physics and Astronomy, Macquarie University, 2109 NSW, Australia }
\affiliation{Istituto Italiano di Tecnologia, Via Morego 30, 16136 Genova, Italy}
\author{Xavier Vidal}
\affiliation{Department of Physics and Astronomy, Macquarie University, 2109 NSW, Australia }
\author{Pawel Wozniak}
\author{Peter Banzer}
\affiliation{Max Planck Institute for the Science of Light, Staudtstr. 2, D-91058 Erlangen, Germany}
\affiliation{Institute of Optics, Information and Photonics, Department of Physics,
Friedrich-Alexander-University Erlangen-Nuremberg, Staudtstr. 7/B2, D-91058 Erlangen, Germany}
\author{Gabriel Molina-Terriza}
\email{gabriel.molina.terriza@gmail.com}
\affiliation{Department of Physics and Astronomy, Macquarie University, 2109 NSW, Australia }

\begin{abstract}
Linear scattering processes are usually described as a function of the parameters of the incident beam. The wavelength, the intensity distribution, the polarization or the phase are among them. Here, we discuss and experimentally demonstrate how the angular momentum and the helicity of light influence the light scattering of spherical particles. We measure the backscattering of a $4\mu m$ TiO$_2$ single particle deposited on a glass substrate. The particle is probed at different wavelengths by different beams with total angular momenta ranging from $-8$ to $+8$ units. It is observed that the spectral behavior of the particle is highly dependent on the angular momentum and helicity of the incoming beam. While some of the properties of the scattered field can be described with a simple resonator model, the scattering of high angular momentum beams requires a deeper understanding of the multipolar modes induced in the sphere. We observe that tailoring these induced multipolar modes can cause a shift and a spectral narrowing of the peaks of the scattering spectrum. Furthermore, specific combinations of helicity and angular momentum for the excitation lead to differences in the conservation of helicity by the system, which has clear consequences on the scattering pattern.
\end{abstract}

\maketitle

\section{Introduction}
The scattering of light by small particles is an important problem in science and engineering. In most situations, an analytical solution to Maxwell equations cannot be found and the problem must be solved computationally. Educated approximations to different particle geometries are used to provide insight into the problem at hand as well as to light-matter interactions in general. There is a notorious exception to this, \textit{i.e.} the scattering of a homogeneous sphere illuminated by a plane wave, which was found to have an analytical solution by Lorenz and Mie in 1898 and 1908 respectively \cite{Lorenz1898,Mie1908}. With the grounds layed out by Lorenz-Mie's theory, a big step was taken in the 1990's when the theory was extended to any kind of incident light beam \cite{Gouesbet1988,Gouesbet2011}. Thanks to these advances, the interaction between lasers beams and particles can be described in a more accurate manner.

Recently, an additional simplification to the scattering of paraxial and non-paraxial beams off spherical particles has been developed using the angular momentum (AM) of light \cite{Zambrana2012,Zambrana2013JQSRT}. The understanding of the AM of light took a big leap forward in the 1990's, with the seminal paper by Allen and co-workers  \cite{Allen1992}. They showed that the AM of light could be controlled using elements such as quarter-wave plates and computer-generated holograms. In the beginning, the AM was especially used in optical tweezers and in quantum optics \cite{Gabi2007,Padgett2011}. More recently, along with the recent developments in vector beams, it has started to be used in nano-optics to control the excitation of nanostructures \cite{Kindler2007,Volpe2009,Banzer2010,Banzer2010OE,Aiello2015,Wozniak2015,Zambrana2016}.  

In this work, we set out to experimentally study the interaction of a micron-sized single sphere with beams with a well-defined AM and helicity. We observe that the AM of light can modify the backscattering and spectrally shift its maxima and minima. Furthermore, we see that the scattering peaks are sharpened thanks to an increase of AM, as predicted in \cite{Zambrana2012,Zambrana2013JQSRT}. Finally, it will be observed that the spectral minima and maxima of the backscattering have very different spatial distributions, and we will relate them to the helicity content of the scattering. We note that a few studies have experimentally investigated the scattering of vortex beams off spheres. In \cite{Garbin2009,Dima2012}, it was shown that the phase singularity of monochromatic vortex beams could be used to exctract information about the position of nanoparticles. In \cite{Kindler2007,Banzer2010,Aiello2015,Wozniak2015}, the authors used vortex beams with polarization singularities to show that the resonant behavior of nanoparticles can be tailored by the excitation beam. Our work differs from those previously reported as it presents a series of light scattering experiments made with beams with a well-defined AM and helicity over a certain spectral range. Overall, we show that the helicity and the AM of light are two crucial properties in order to understand, predict and control the light scattering of particles. 

\section{Results}

\subsection{Experimental set-up}
\begin{figure}[tbp]
\centering
\includegraphics[width=14cm]{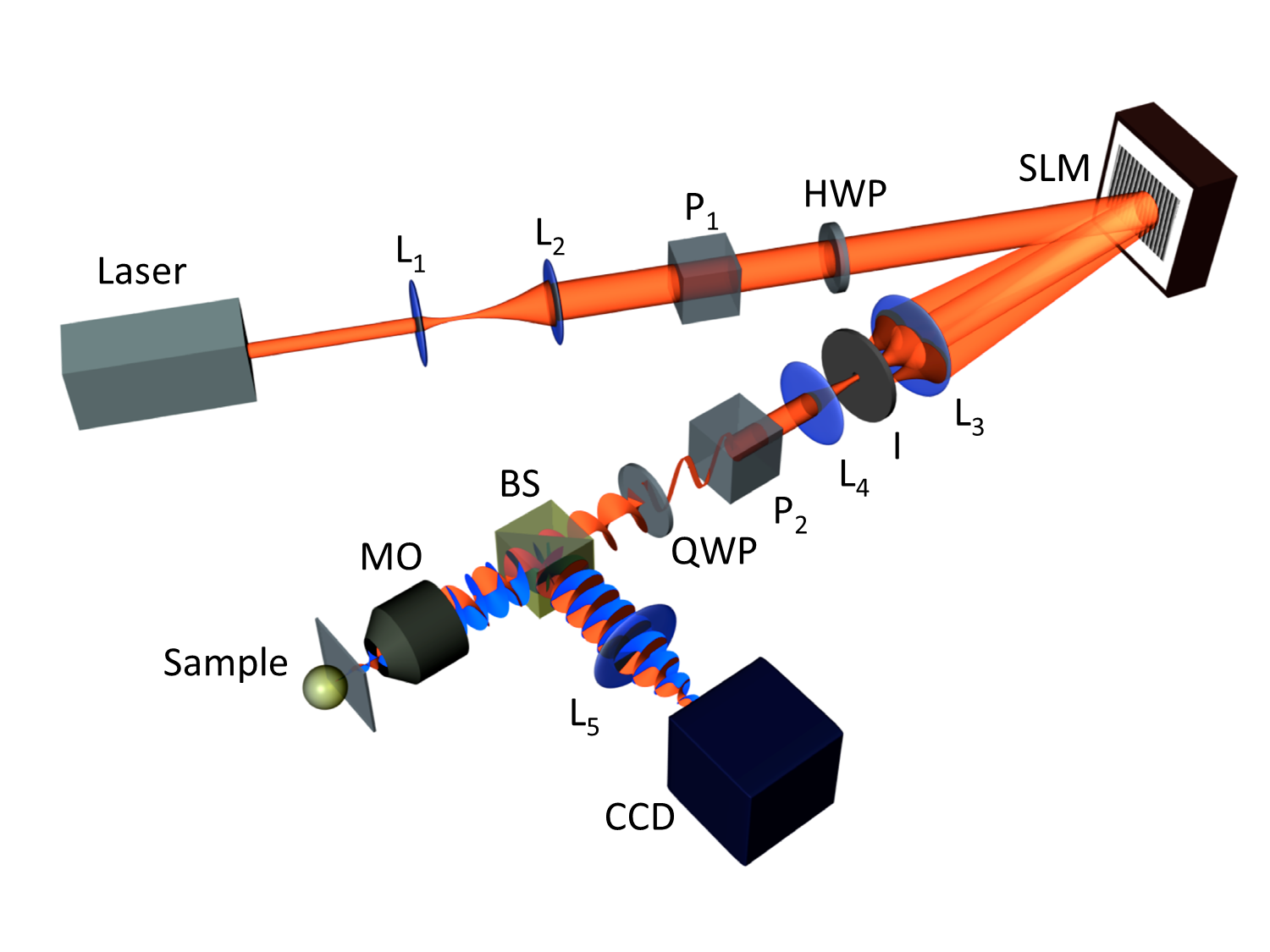} 
\caption{Schematics of the experimental set-up. A laser beam of variable wavelength $\lambda=[760-810]$ nm is expanded with two lenses (L$_1$, L$_2$) and polarized with a linear polariser (P$_1$) and a half-wave plate (HWP) to match the polarization axis of the SLM. The SLM gives the desired spiral phase to the beam, and the first diffraction order is selected using two lenses (L$_3$, L$_4$) and an iris (I). The two lenses are simultaneously used to expand the beam and match the size of the back-aperture of the microscope objective (MO). The polarisation of the beam is modified with a linear polariser (P$_2$) and a quarter-wave plate (QWP). The beam is focused onto the sample using a MO and the backscattering is collected with the same MO, and then separated from the incident beam by a beam splitter (BS). The backscattered light is imaged onto a CCD camera using L$_5$, hence imaging the focal plane. \label{fig:sch}}
\end{figure}

The experimental set-up is schematically displayed in Figure \ref{fig:sch}. A tunable fiber diode laser yielding a Gaussian mode is lineraly polarized and it illuminates the chip of a spatial light modulator (SLM). Afterwards, the SLM modifies the phase of the beam to create a vortex beam with a phase singularity of order $l$. The first diffraction order is selected using a pair of lenses and an iris (L$_3$, L$_4$ and I). Then, its polarization is modified to create a state of well-defined helicity with a linear polariser and a quarter-wave plate (P$_2$ and QWP) \cite{Nora2014,Zambrana2016}. Using a water-immersion microscope objective (MO) with a numerical aperture of $\mathrm{NA}=1.1$, the light beam is focused onto the sample. The exit pupil of MO is completely filled, yielding a spot size with a diameter of 400nm approximately.

The sample consists of a set of single TiO$_2$ (amorphous crystal phase) particles deposited on top of a microscope slide. The separation of the TiO$_2$ particles is not constant. To avoid couplings between neighboring particles, we have only probed a particle whose closest neighbors are more than 20$\mu m$ apart. The titania particles have a radius of $R=2\mu m$, and their index of refraction in air is $n_r=1.8$. The slide is placed on a sample holder, which is attached to a nano-positioning device. Note that MO is a long working distance water immersion objective, therefore the incoming beam goes through water and the microscope slide first, and then it interacts with the single particle. The center of the particle is placed on the optical axis and at the focal plane of the incoming beam.

The same MO used to focus the light on the sample is used to collect the backscattering. A beam-splitter (BS) is used to separate the backscattered light from the incident beam.  Afterwards, the reflected beam passes through a lens (L$_5$) that images the sample onto a CCD camera.

\subsection{Measurements details}
In this experiment, we want to probe the effects that the AM of light has on the spectral response of a single dielectric sphere. For this purpose, we use large particles ($R=2\mu m$) and an SLM. The reason why micron-sized particles are needed will be analytically discussed later. The light beams will be described using three well-defined properties - their wavelength ($\lambda$), their helicity ($p$) and their AM along the direction of propagation of the beam ($m_z$). Due to the existing mirror symmetries between these types of beams \cite{Zambrana2014Nat,Zambrana2016}, we do our experiments with the beams with helicity $p=1$, and we will use the other helicity $p=-1$ as a control, \textit{i.e.} to double-check the validity of the results. The mirror symmetry guarantees that the backscattering of a beam of light characterized by $\left(  \lambda, m_z, p\right)$ will be equal to another beam with $\left(  \lambda,- m_z, -p\right)$ \cite{Zambrana2014Nat,Zambrana2016}. Note that the description of the beams holds both for the paraxial and the non-paraxial regime, since an aplanatic lens does not modify the value of any of these three parameters \cite{Ivan2012PRA,Nora2014,Zambrana2016}. As for the effect of the glass substrate on the beam description, it has no effect on $\lambda$ and $m_z$, and its effect on $p$ can be neglected \cite{Nora2014}. 

Experimentally, the key elements to obtain beams of well-defined AM and helicity are the SLM and the QWP (see Figure \ref{fig:sch}). The SLM is used to create a vortex beam with a phase singularity of order $l$. The QWP is used to create a state of circular polarization with one of the two possible handedness. Because the beam is paraxial, we can approximately consider that if the beam is left circularly polarized, its helicity is $p=1$ \cite{Ivan2012PRA}. Hence, a paraxial right circularly polarized beam is described with $p=-1$. Note that it can be demonstrated that a collimated beam with helicity $p$ and a phase singularity of order $l$ along its axis of propagation has a total AM along the same axis of $m_z=l+p$ \cite{Nora2014,Zambrana2014Nat,Zambrana2016}. 

For our measurements, we have used vortex beams with phase singularities of the order $l=0, \pm 1, \pm 2, \pm 6, \pm 7$. The wavelength of the monochromatic laser has been tuned from 760 to 810nm in steps of $\Delta\lambda=4$nm. This spectral range has been chosen as the particle supports higher order resonances in this range. 

\subsection{Experimental results}
\begin{figure}[tbp]
\centering
\begin{subfigure}{.5\textwidth}
  \centering
  \includegraphics[width=8.2cm]{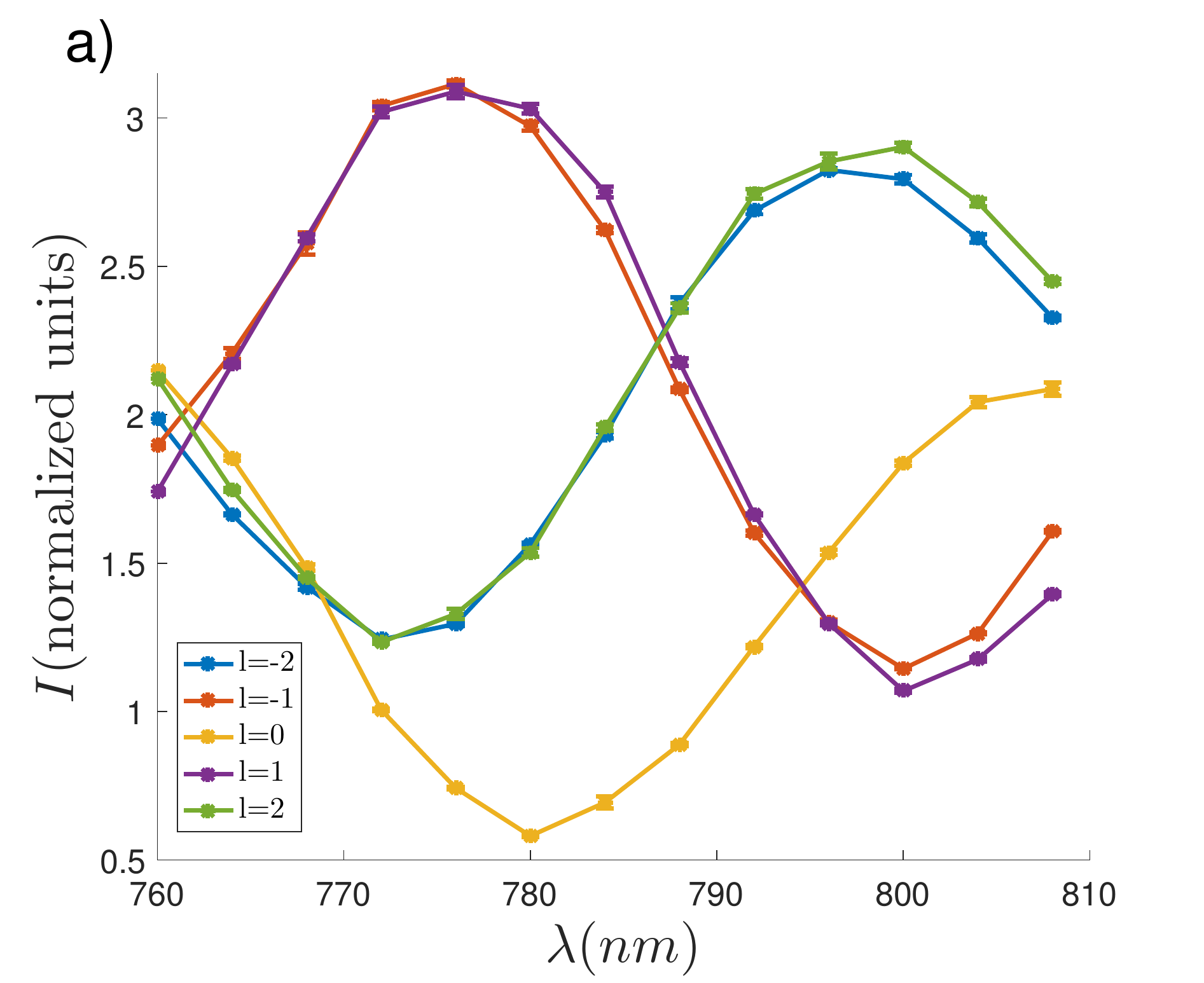}
  \label{fig:subL0}
\end{subfigure}%
\begin{subfigure}{.5\textwidth}
  \centering
  \includegraphics[width=8.2cm]{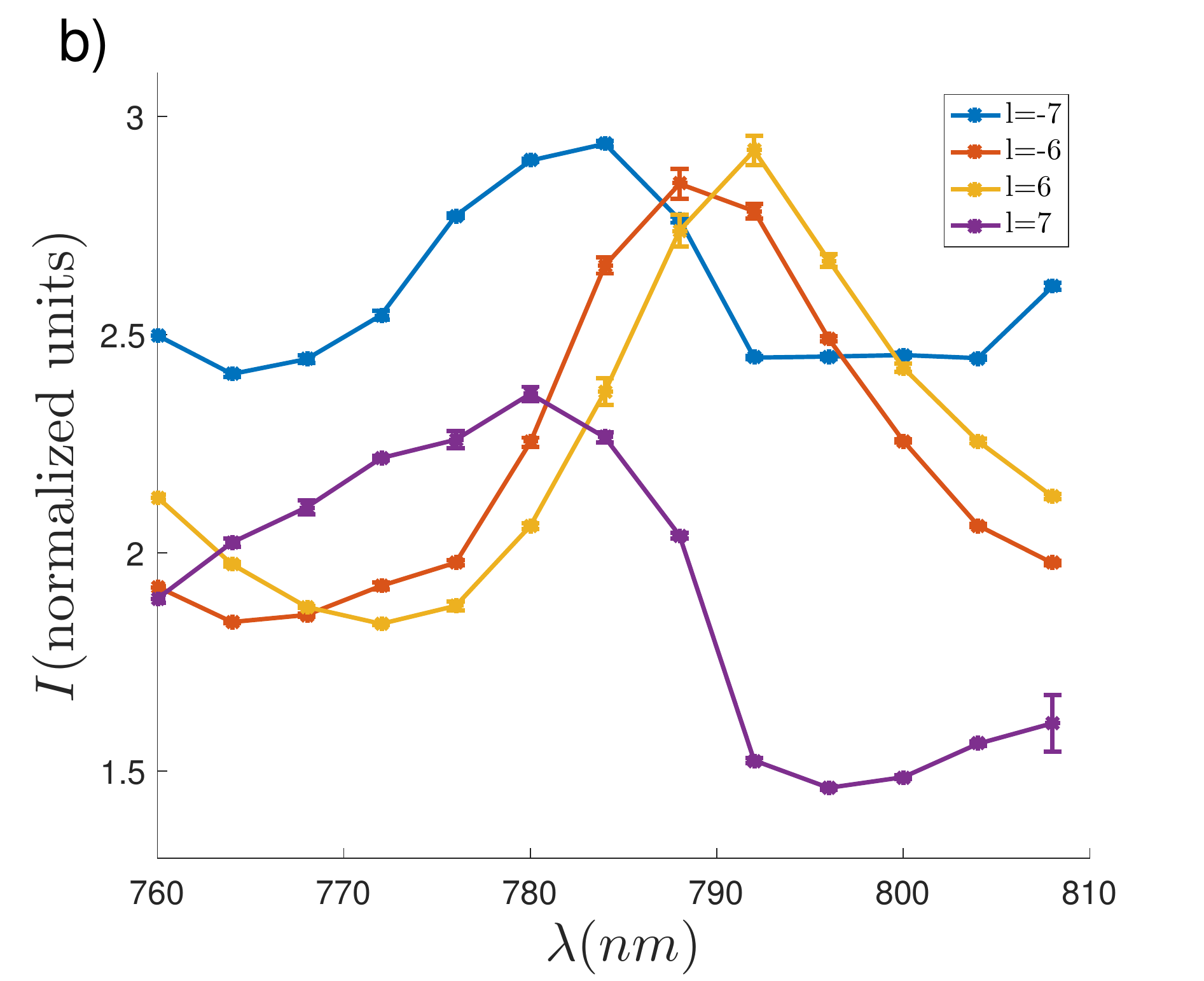}
  \label{fig:subL7}
\end{subfigure}
\caption{Experimental backscattering spectra of a 4 $\mu$m TiO$_2$ particle placed on top of a microscope slide. The particle is probed with beams of helicity $p=1$ and $\lambda=760-810$ nm in 4 nm steps. The excitation beams have a) $l=0,\pm 1, \pm2$ and b) $l=\pm 6, \pm 7$. The signal is normalized to the power of reflected beams at the glass-air interface.  }
\label{fig:scan_l}
\end{figure}
\begin{figure}[tbp]
\centering
\includegraphics[width=14cm]{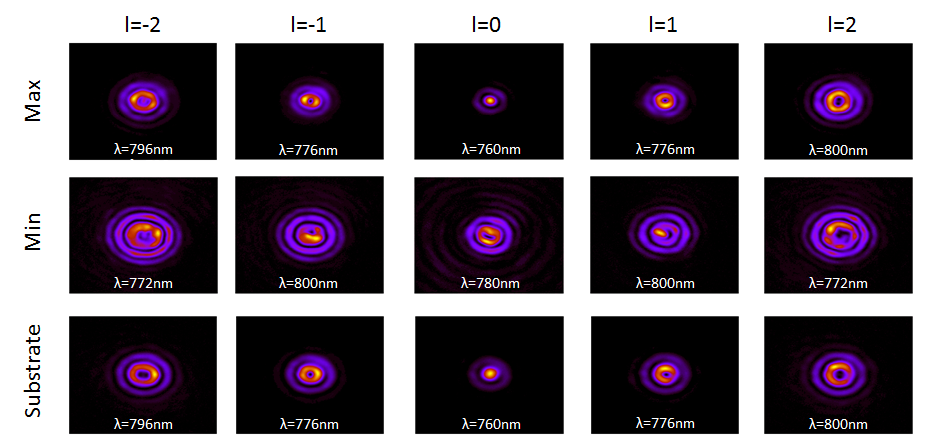} 
\caption{Snapshots of the TiO$_2$ particle's backscattering for incident beams with phase singularities of the order $l=0,\pm 1,\pm 2$. The snapshots for the particle's backscattering maxima and the minima are taken at their corresponding wavelength, as observed in Fig. \ref{fig:scan_r}. The snapshots of the substrate were taken at the wavelength of the maximum backscattering. The corresponding wavelengths are labelled at the bottom of each snapshot. \label{fig:beams}}
\end{figure}
\begin{figure}[tbp]
\centering
\begin{subfigure}{.5\textwidth}
  \centering
  \includegraphics[width=8.2cm]{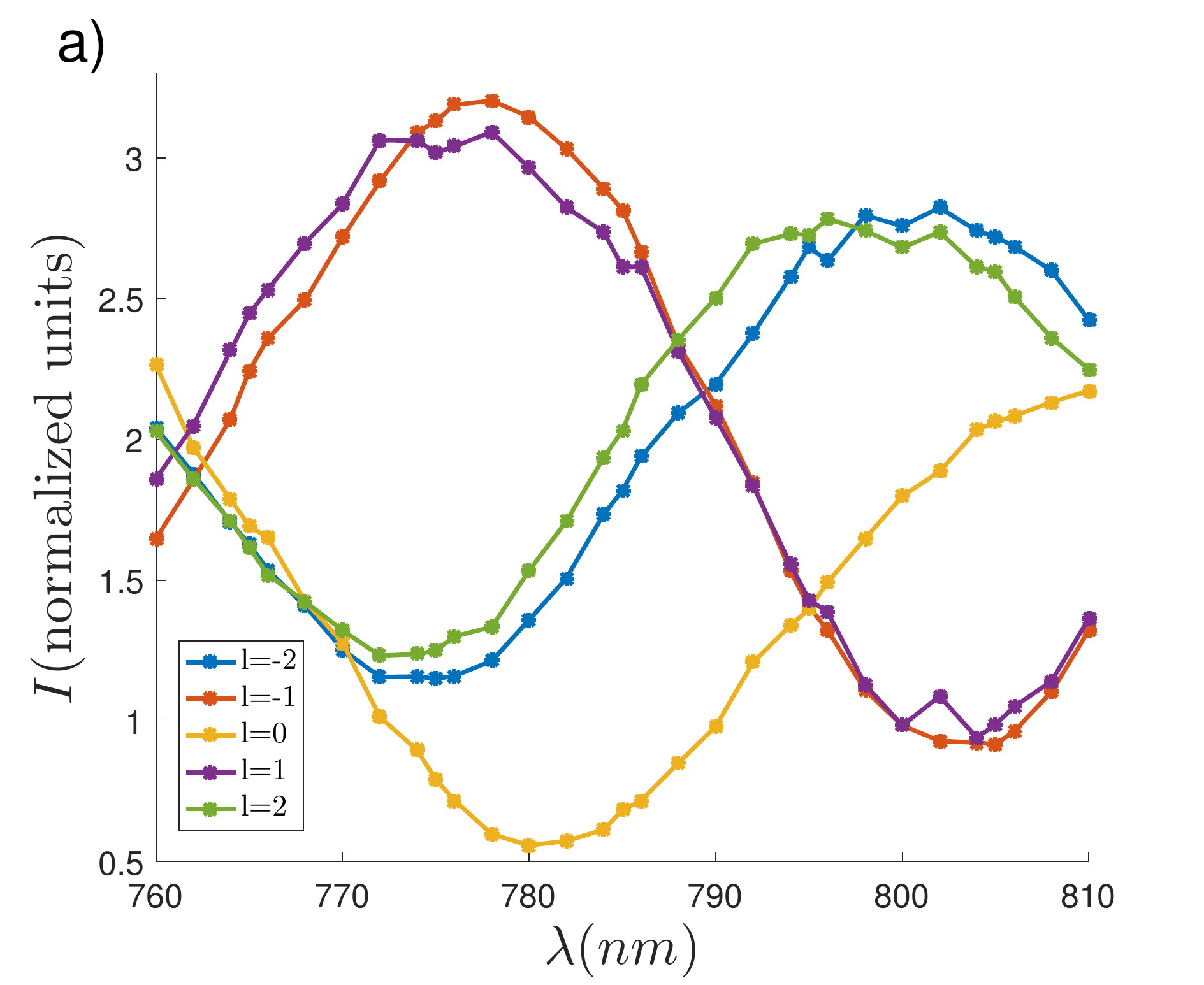}
\end{subfigure}%
\begin{subfigure}{.5\textwidth}
  \centering
  \includegraphics[width=8.2cm]{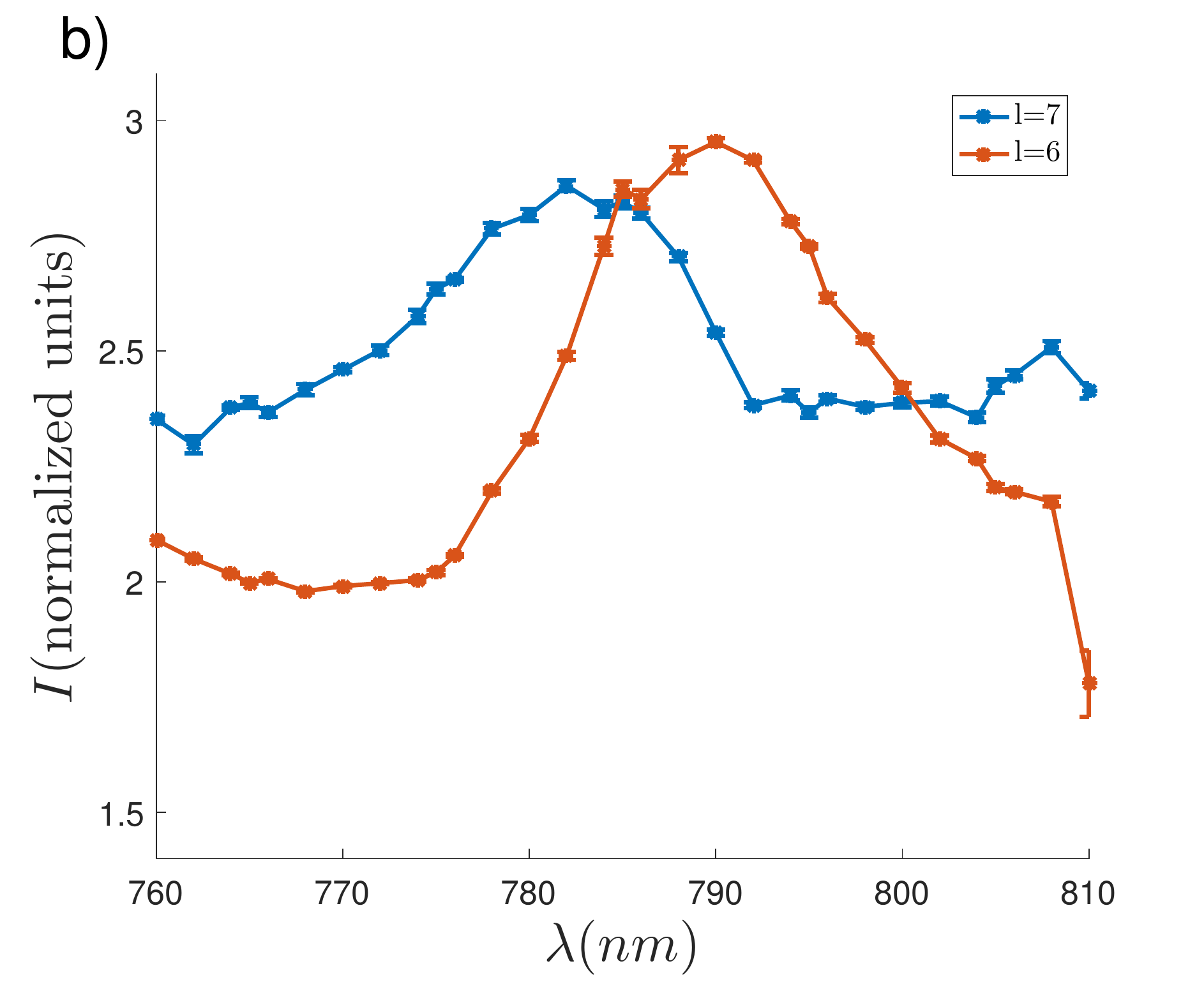}
\end{subfigure}
\caption{Backscattering wavelength scans for excitation beams of helicity $p=-1$ and $\lambda=760-810$ nm with a step size of 2 nm. a) The excitation beams have $l=0,\pm 1, \pm2$. b) The excitation beams have $l= 6, 7$. }
\label{fig:scan_r}
\end{figure}

In Figures \ref{fig:scan_l}-\ref{fig:scan_r} we show the collected experimental data. In Figure \ref{fig:scan_l} we plot the backscattering spectra of the TiO$_2$ particle for $\lambda=760-810$nm. The backscattering is given as a  dimensionless number $I$. The value of $I$ is computed as the ratio between the reflection of the particle when it is placed on the optical axis of the beam, over the reflection from the glass-air interface when the particle has been displaced 10 $\mu$m off the optical axis. The reflection signal is obtained integrating a restricted area in the snapshots taken by the CCD camera (see Methods). In order to make the plots clearer, we have divided the data in two subfigures, a) for $l=0, \pm 1, \pm 2$ and b) for $l=\pm 6, \pm 7$. The helicity of the beam is $p=1$ for both cases. 

In Figure \ref{fig:scan_l}a) we observe that the backscattering has a spectral oscillating behavior for all the incident vortex beams. These oscillations as a function of the wavelength have also been observed in a similar configuration in \cite{Ashkin1981,Barton1989a}. However, to the best of our knowledge, the dependence of these oscillations on the excitation beam has never been discussed previously. It is observed that a circularly polarized Gaussian beam ($l=0$) yields a minimum of backscattering at $\lambda=780$nm. In contrast, the beams with $l=\pm 1, \pm 2$ have both a maximum and a minimum in the probed range. We see that the beams of the same value $\vert l \vert$ yield also the same backscattering spectra. 

In Figure \ref{fig:scan_l}b), we see observe a different behavior. Firstly, the beams with the same $\vert l \vert$ do not yield the same result - even though they still resemble each other. Furthermore, in comparison to Figure \ref{fig:scan_l}a), we observe that the oscillating behavior is shifted and sharpened. While the spectral features for low-order angular momentum modes do not depend on the sign of $l$, the backscattering changes with the sign of $l$ for high angular momenta modes.

To gain a better understanding of the back-reflection off spherical particles, we shall analyze its intensity distribution with a CCD camera. The images were integrated to obtain the data displayed in Figure \ref{fig:scan_l} (see Methods), but they also provide useful insight on the features of the scattering. Figure \ref{fig:beams} shows the recorded images arranged in five columns labeled with the azimuthal phase $l$ imprinted by the SLM onto the beam used to produce the images. Each column shows the backscattering obtained with an incident beam with $p=1$ and $m_z=l+p$. The two upper rows show the signal at the CCD camera for different wavelengths, namely the maxima and minima obtained in Figure \ref{fig:scan_l}. The third row shows the backscattering pattern of the substrate, \textit{i.e.} it has obtained when the particle was moved out of the reach of the incident beam. Even though the substrate images are taken at the wavelength corresponding to the maxima (first row), it has been verified that the pattern is almost identical in the chosen studied spectral range. Now, by comparing the two upper rows, it is observed that the spatial distributions of the backscattered field vary significantly even in a short wavelength range. Moreover, we observe that the images depicted in the two rows corresponding to maximum backscattering and substrate only are almost identical.  

Finally, in Figure \ref{fig:scan_r} we show the results obtained for an incident beam with helicity $p=-1$ and $l=0, \pm 1, \pm 2, 6, 7$. As mentioned before, this was done to cross-check the obtained experimental data. It is observed that the results are mirror symmetric, as it was expected \cite{Zambrana2014Nat,ZambranaThesis,Zambrana2016}. Indeed, the spectra exhibit the same spectral behavior as well as comparable quantitative values. Also, notice that the wavelength scans for $p=-1$ have been done with a $\Delta\lambda<4$nm.

\section{Discussion}
In the previous section we have experimentally shown that the angular momentum of a beam can substantially change the scattering properties of a particle. Let us discuss these results and their relevance to light matter interaction in general. First of all, Figure \ref{fig:scan_l} and its mirror symmetric counterpart Figure \ref{fig:scan_r} show us that the strength of the scattering process strongly depends on the wavelength and it has an oscillating behavior. Furthermore, it is seen that the oscillating behavior can be shifted and narrowed depending on the properties of the incoming beam. For Gaussian beams, similar oscillations, as well as the so-called ripple structure, were measured and described in \cite{Ashkin1981}. There are many ways of explaining the appearance of this kind of oscillations, depending on the level of approximation used. Ashkin \textit{et al.} explained that, given that most of the light is concentrated in the center of the beam and the light is tightly focused, one could consider this effect as a simple lossy resonator \cite{Ashkin1981}. Hence, an oscillating behavior is expected. The model of a lossy resonator predicts that there is a countably infinite number of resonant wavelengths \cite{Saleh1991}. These maxima are located at $\lambda_N = 2Ln/N$, where $n$ is the refractive index of the resonator, $L$ is the length of the ``cavity'' and $N$ is a positive integer that numerates the position of the maxima \cite{Saleh1991}. Here, the comparison is made with the minima of Figure \ref{fig:scan_l}a), as the minima of backscattering correspond to the resonant condition of minimum leakage. We can see that, given $n=1.8$ and $L=2R= 4\mu$m, the oscillations in Figure \ref{fig:scan_l}a) are predicted relatively well by $N=18$ and $N=19$. For instance, we can see that the minimum of $l=\pm1$ are at 800 nm = $2 \cdot 4$ $\mu$m $\cdot 1.8 / 18$. Moreover, the spectral distance between $\lambda_{18}$ and $\lambda_{19}$ is $2Ln(\frac{1}{18}-\frac{1}{19})=42$nm, which is approximately the distance between two consecutive minima for all the curves in Figure \ref{fig:scan_l}a). Higher order modes seem to depart from the scenario depicted by Ashkin and coworkers, but it is possible to push this picture even further. Given the different Guoy phase shifts of optical modes, it is expected that there is a change in the oscillatory behaviour for even and odd $l$ numbers \cite{Saleh1991}. This effect is observed for beams with $l=\pm1, \pm2$ in Figures \ref{fig:scan_l}a) and \ref{fig:scan_r}a), where the spectral maxima of the modes with $l=\pm 1$ are indeed at the minima of $l=\pm 2$. However, this simple representation fails to describe the scattering of greater angular momentum modes, and we have to resort to a full development of the scattering problem.

A description of a similar effect was presented in \cite{Zambrana2012,Orlov2012}. There, it is shown that cylindrically symmetric beams can be easily decomposed into electromagnetic multipolar modes. The semianalytical decomposition yields two simple rules. 1) The more tightly a beam is focused, the fewer multipolar modes its decomposition contains - with the opposite case being that of the plane wave, which populates all multipolar modes. 2) The multipolar decomposition of a beam with angular momentum $m_z$ only contains multipolar modes whose order is $j \geq m_z$. Now, light scattering from particles is always given as a superposition of modes. In general, the number of multipoles that take part in the scattering process grows with the size of the particle. Besides, it is known the Q-factor of Mie resonances increases with the order $j$ of the resonance \cite{Oraevsky2002}. Therefore, the scattering cross section features sharper resonances (higher Q-factors) when the scattering is dominated by higher-order modes. This is one of the reasons why spherical whispering gallery mode resonators are at least 20 $\mu m$ in size, so that resonances of the order $j=2000$ can be excited \cite{Oraevsky2002}. From this perspective, it is important to note that: i) Tightly focused beams with a low AM will produce broader oscillations of the backscattering, as lower multipolar modes will dominate the scattering \cite{Zambrana2012}. ii) When we increase the angular momentum of the beams, the lower order modes will play a much less relevant role in the superposition, and scattering cross sections with sharper resonances will be obtained \cite{Zambrana2012,Zambrana2013JQSRT}.


Quantitatively, this can be formalized as follows. According to Mie theory, the field scattered off a particle is analytically found once the illumination has been decomposed into multipolar modes \cite{Zambrana2012,Zambrana2013JQSRT,ZambranaThesis}. That is, if the incident beam is expressed as $\Ei = \displaystyle\sum_{j,m_z} \aje \Ae + \ajm \Am$, the scattered field is:
\begin{equation}
\Es = \displaystyle\sum_{j,m_z} \left[ \aje a_j \Ae + \ajm b_j \Am \right]
\end{equation}
where $a_j,b_j$ are the so-called Mie coefficients, $\Ae,\Am$ are the electric and magnetic multipolar modes, and $\aje, \ajm$ are general multipolar coefficients modulating the superposition. Due to the orthonormality relations between $\Ae$ and $\Am$, the scattering cross section is computed as:
\begin{equation}
C_s \propto \displaystyle\sum_{j=1}^{\infty}\sum_{m_z=-j}^{j} \left( \vert \aje a_j \vert^2 + \vert \ajm b_j \vert^2 \right)
\label{eq:Cs}
\end{equation}
As mentioned before, we can define the vortex beams used in the experiment as a function of $\lambda,p,m_z$. Hence, we will denote $\Emp$ as the incident monochromatic beam with helicity $p$, and AM $m_z$. In \cite{Zambrana2012,Zambrana2013JQSRT}, it was demonstrated that if $\Ei = \Emp$, then
\begin{equation}
C_s \propto \displaystyle\sum_{j=\vert m_z \vert}^{\infty} \vert C_{jm_z} \vert^2 \left( \vert a_j \vert^2 + \vert  b_j \vert^2 \right)
\label{eq:Cs_cyl}
\end{equation}
where $C_{jm_z}$ is the only set of coefficients that define the multipolar content of the incident beam. We shall use eq.(\ref{eq:Cs_cyl}) to discuss the experimental results. 

Notice that the coefficients $\Cjm$ (same applies to $\aje,\ajm$) have a very smooth behavior as a function of the wavelength. Therefore, the wavelength dependence seen in Figures \ref{fig:scan_l} and \ref{fig:scan_r} needs to come from the Mie coefficients $a_j,b_j$. Moreover, when we change the value of the AM of the incoming beam (we change $m_z$), we modify the function $\Cjm$ and, hence the number of the excited modes of the particle which contribute to the scattering process. That is, a given $m_z$ of the incident beam ensures that the contribution of the multipolar modes whose order is lower than $j=\left( \vert m_z \vert -1 \right) $ is suppressed from the scattering \cite{Zambrana2012,Zambrana2013JQSRT,Zambrana2013OE}. Here we can distinguish between three scenarios depending on the AM of the incident beam. If $m_z \gtrsim 2\pi n_r R/\lambda$, the scattering of the particle is completely suppressed \footnote{Indeed, it can be checked that the Mie coefficients $a_j$ and $b_j$ are negligible when $j=m_z \gtrsim 2\pi n_r R/\lambda$.}. If $m_z \sim 2\pi n_r R/\lambda$, a single (or very few) multipolar mode can be excited \cite{Zambrana2012,Zambrana2013JQSRT,Zambrana2013OE}. And if $m_z \lesssim 2\pi n_r R/\lambda$, which is the case of our experiment, different modes contribute to the scattering. Note that this the reason why micron-sized particles have been used, so that the scattering is not entirely suppressed when using high AM modes. Given a particle defined by ($R, n_r$) and an excitation wavelength $\lambda$, the lowest $j \approx \left( 2\pi n_r R/\lambda \right)$ multipolar modes will potentially contribute to the scattering. By increasing the AM content of the incident beam ($m_z$), the contribution of the lowest $j=\left( \vert m_z \vert -1 \right)$ multipolar  modes is progressively suppressed, until there are no contributing modes. Since higher Mie resonances are narrower, it follows that when the low-order modes are suppressed from scattering, the spectral response of the system is sharper. This effect was theoretically shown in \cite{Zambrana2012,Zambrana2013JQSRT}, and here we experimentally observe it in Figures \ref{fig:scan_l} and \ref{fig:scan_r}. Clearly, the oscillations depicted in Figures \ref{fig:scan_l}a) and \ref{fig:scan_r}a) are broader than those shown in Figures \ref{fig:scan_l}b) and \ref{fig:scan_r}b). 

As for the spectral shift of the oscillations in Figures \ref{fig:scan_l} and \ref{fig:scan_r}, it is a subtle effect due to the superposition of different modes. In Methods, we show the results of some numerical simulations using a single TiO$_2$ sphere embedded in air with $R=2\mu m$ and $n_r=1.8$. The simulations show that the backscattering spectral response never resembles that of a single multipolar mode. Instead, the response of the particle is always a superposition of at least five multipolar modes. This feature highlights the fact that the spectral measurements shown in Figures \ref{fig:scan_l} and \ref{fig:scan_r} are oscillations, and not Mie resonances. We make a clear distinction between scattering oscillations and Mie resonances. On the one hand, Mie resonances are the eigenresonances of a sphere, and thererefore they cannot be spectrally tailored with an incoming beam. Their spectral value is fixed by the geometry and material properties of the particle \cite{Zambrana2015}. On the other hand, scattering oscillations are the product of a complex superposition of Mie modes. These complex superpositions give rise to local scattering minima and maxima, and these are the ones that we define as oscillations. When we change the incoming beam, we change the complex superposition of modes that give rise to oscillations, and hence their spectral position can vary. In contrast, Mie resonances are always located in the same spectral position regardless of the incoming beam. 

Finally, we shall explain the differences in the intensity distributions of backscattered light depicted in Figure \ref{fig:beams}. The images show the backscattered light of the system formed by the particle and the glass-air interface excited with a beam that can be modelled as $\Ei=\Emp$. Here, as in the rest of the paper, backscattering needs to be understood as the light scattered in the reflection semi-space (see Figure \ref{fig:sch}). The scattering of a system can be divided into its two helicity component \cite{Ivan2012PRA,Nora2014,Zambrana2016,Zambrana2016Nano}, \textit{i.e.} $\Es=\mathbf{E}_{p}^{\mathbf{s}}+\mathbf{E}_{-p}^{\mathbf{s}}$, where the two components have a well-defined helicity \cite{Zambrana2016}. Next, we are going to argue that the differences in the images stem from the relative difference between these two helicity components. But before, a point needs to be made about the reflection of circularly polarized light by a glass-air planar interface.

As it was commented in the Results section, the substrate by itself yields a very constant response as a function of the wavelength. It behaves as a mirror with a very low reflectivity. That is, the backscattered light off the substrate is a mirror image of the incident beam. Now, the mirror image of a beam $\Emp$ is $\mathbf{E}_{-p,m_z}^{\lambda}$. Notice that a beam like $\mathbf{E}_{-p,m_z}^{\lambda}$ in the backward semi-space yields the same intensity pattern as $\Emp$ in the forward semi-space \cite{ZambranaThesis}. That is why the third row of Figure \ref{fig:beams} yields an approximate mirror image of the incident beam $\Emp$.

\begin{figure}[tbp]
\centering
\includegraphics[width=14cm]{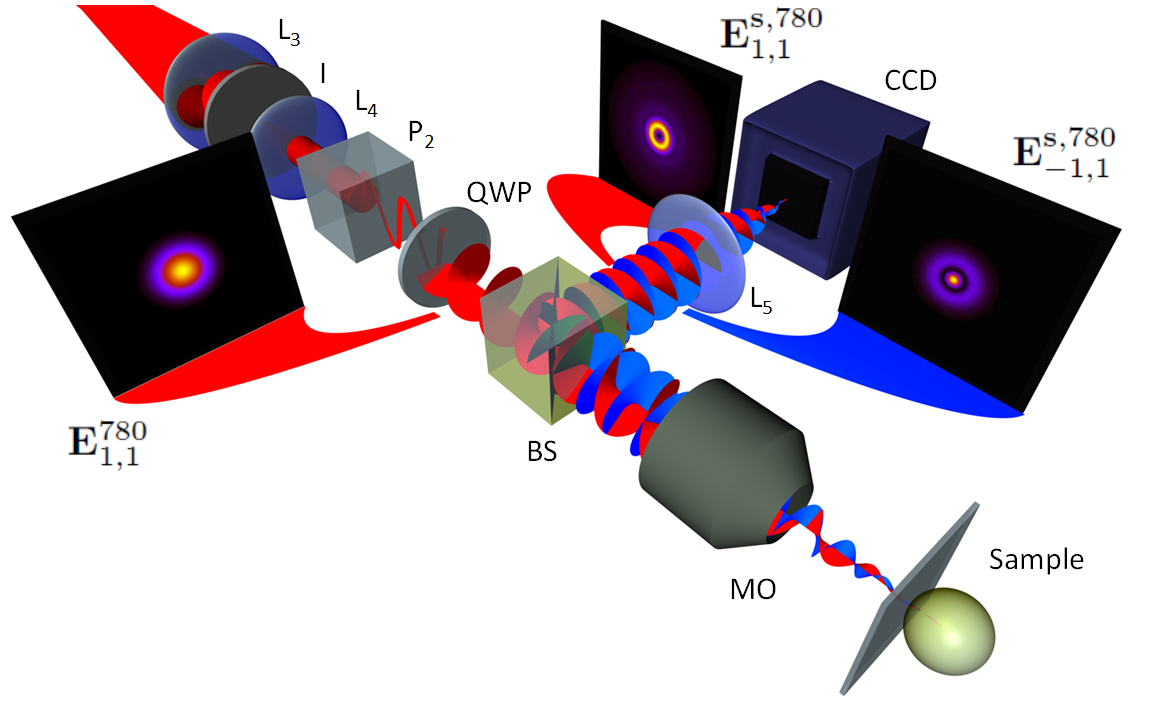} 
\caption{Schematics of a backscattering event decomposed into its two helicity components. A circularly polarized Gaussian beam $\Ei=\mathbf{E}_{1,1}^{780}$ is prepared using the apparatus depicted in Figure \ref{fig:sch}. Its intensity distribution is plotted on the left hand side of the picture. The incident beam is focused onto a TiO$_2$ which lays on top a glass substrate. The backscattering is collected with the same MO and it is splitted from the incident light with a BS. The backscattering is schematically decomposed into its two helicity components - $\mathbf{E}_{1,1}^{\mathbf{s},780}$ and $\mathbf{E}_{-1,1}^{\mathbf{s},780}$. The backscattered helicity component with the same helicity as the incident beam ($\mathbf{E}_{1,1}^{\mathbf{s},780}$, red) is plotted on the left, and the component with the opposite helicity ($\mathbf{E}_{-1,1}^{\mathbf{s},780}$, blue) is plotted on the right. The intensity distribution of $\mathbf{E}_{1,1}^{\mathbf{s},780}$ contains a phase singularity of charge $l=2$, whereas the intensity distribution $\mathbf{E}_{-1,1}^{\mathbf{s},780}$ does not contain any phase singularity.\label{fig:hel_comp}}
\end{figure}

In general, we denote the scattered field off the system as $\Es$. Since the system scatters light elastically and it is cylindrically symmetric, the system must conserve the wavelength $\lambda$ and the AM of the incident beam, $m_z$. Hence, $\Es \equiv \mathbf{E}_{m_z}^{\mathbf{s},\lambda}$. In contrast, the helicity of light of the incident beam $p$ is not conserved in this scattering process \cite{Ivan2013,Nora2014,Zambrana2016}, therefore the scattered field can be decomposed as $\mathbf{E}_{p,m_z}^{\mathbf{s},\lambda}+\mathbf{E}_{-p,m_z}^{\mathbf{s},\lambda}$. As we depict in Figure \ref{fig:hel_comp}, these two components have a very different spatial dependence. If the incident beam is a circularly polarized Gaussian beam (vortex with $l=0$) $\Ei=\Emp=\mathbf{E}_{1,1}^{780}$, the two backscattered helicity components are $\mathbf{E}_{-1,1}^{\mathbf{s},780}$, and $\mathbf{E}_{1,1}^{\mathbf{s},780}$ (see Figure \ref{fig:hel_comp}). We observe that the intensity distribution of $\mathbf{E}_{-1,1}^{\mathbf{s},780}$ resembles the reflection of the substrate for $l=0$ (see Figure \ref{fig:beams}). However, the other helicity component $\mathbf{E}_{1,1}^{\mathbf{s},780}$ resembles that of a vortex beam with $l=2$. And this is exactly what is observed in Figure \ref{fig:beams} at the minimum of the backscattering plots for $l=0$ - except for some experimental imperfections, which split the phase singularity of charge $l=+2$ into two singularities of charge $l=+1$ \cite{Rich2014,Zambrana2016}. Thus, it seems as if the minima of the oscillations in the backscattering spectral measurements were related to different relative contributions of the two helicity components.

\begin{figure}[tbp]
\centering
\includegraphics[width=14cm]{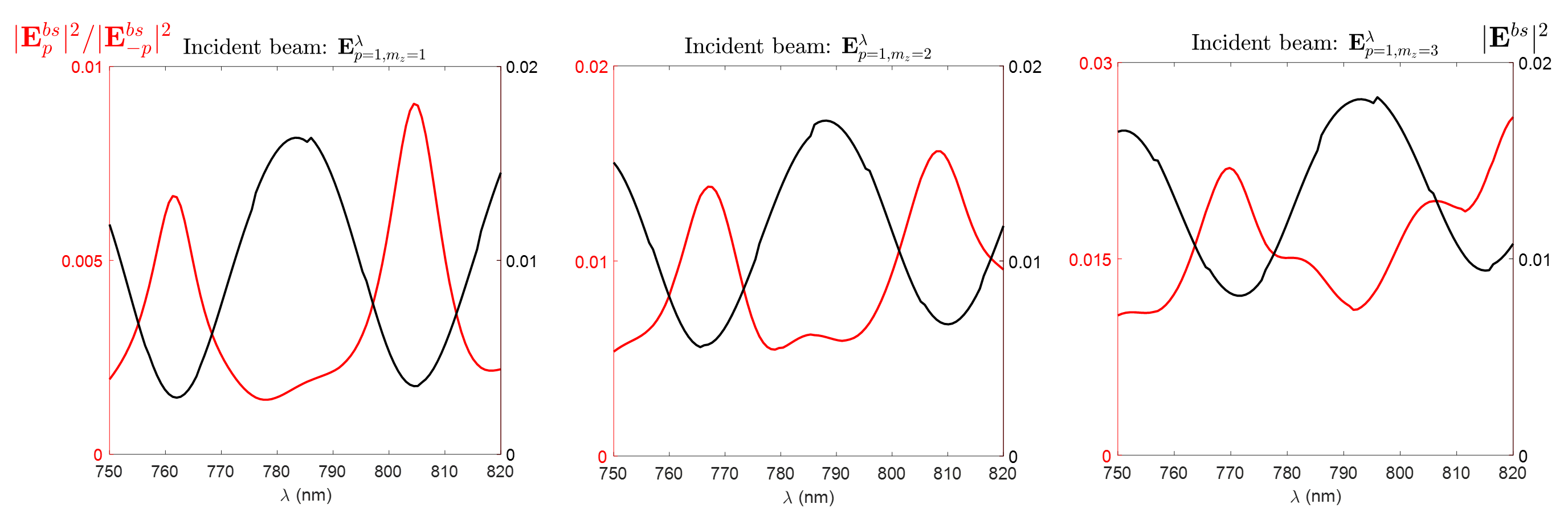} 
\caption{In red (left axis), plots of the helicity conservation computed as a squared ratio of the two backscattered helicity components $\vert \mathbf{E}_{p,m_z}^{\mathbf{bs},\lambda} \vert^2 / \vert \mathbf{E}_{-p,m_z}^{\mathbf{bs},\lambda} \vert^2$. In black (right axis), plots of the backscattered cross section. From left to right, the plots have been obtained with $ \mathbf{E}_{1,1}^{\lambda} $, $\mathbf{E}_{1,2}^{\lambda}$ and $\mathbf{E}_{1,3}^{\lambda}$ incident beams respectively.  All simulations are done in a Mie theory regime, \textit{i.e.} the substrate has not been taken into account.   \label{fig:dual}}
\end{figure}

Next, using Mie theory, we simulate the two backscattering helicity components and we observe their relative contributions. This is displayed in Figure \ref{fig:dual}, where the black curves show the backscattering (see Methods) and the red curves show the ratio between the two helicity squared components. Indeed, the minima in the back-scattered light correspond to approximately the maxima in the $\vert \mathbf{E}_{p,m_z}^{\mathbf{bs},\lambda} \vert^2 / \vert \mathbf{E}_{-p,m_z}^{\mathbf{bs},\lambda} \vert^2$ ratio - with $ \mathbf{E}^{\mathbf{bs}}$ being the scattered field in the backward semi-space. That is, the minima of the backscattering are related to maxima in helicity conservation. This is consistent with the so-called first Kerker condition \cite{Kerker1983,Gomezmedina11,Garcia-Camara11,Aitzol2011}, which explains the suppression of backscattering under certain conditions. As it was shown in \cite{Zambrana2013,Ivan2013}, the first Kerker condition is the consequence of helicity and AM conservation. 

\begin{figure}[tbp]
\centering
\includegraphics[width=14cm]{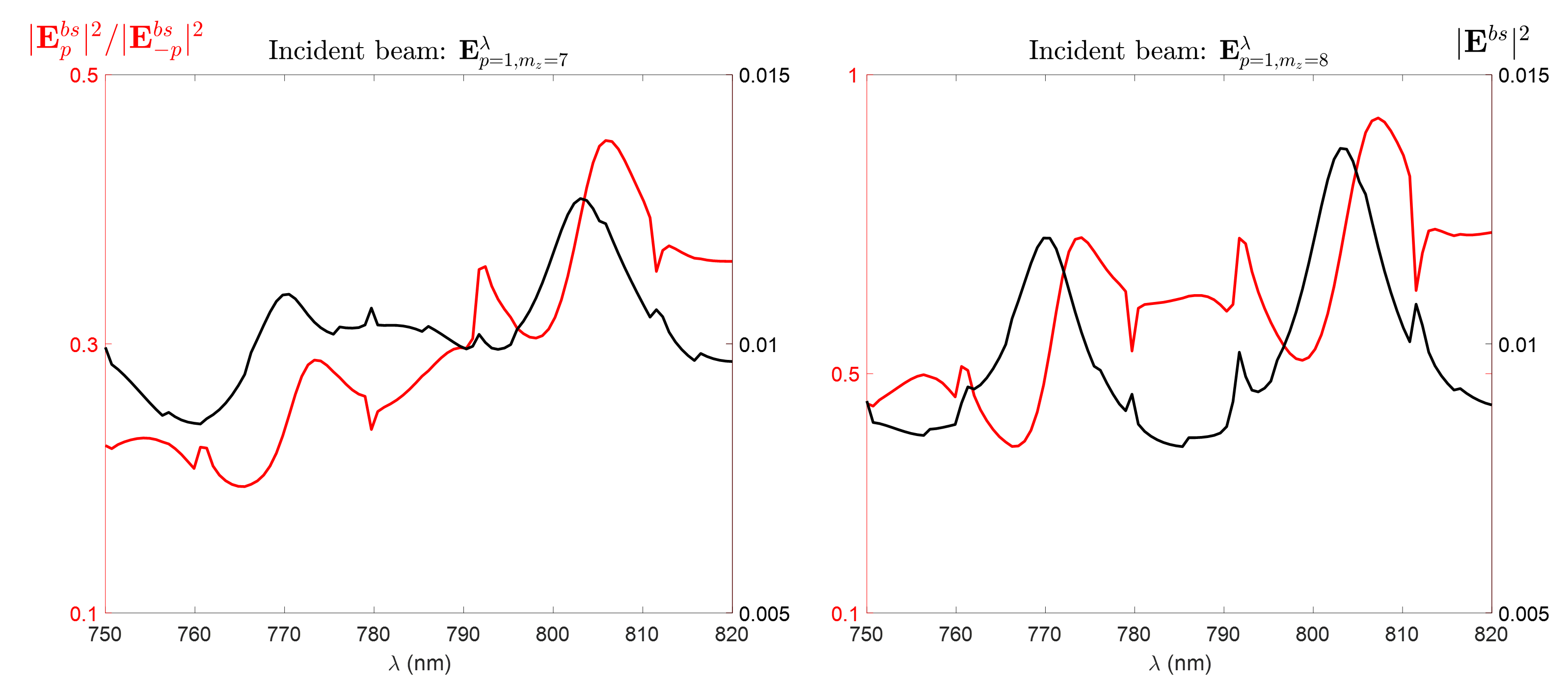} 
\caption{In red (left axis), plots of the helicity conservation computed as a squared ratio of the two backscattered helicity components $\vert \mathbf{E}_{p,m_z}^{\mathbf{bs},\lambda} \vert^2 / \vert \mathbf{E}_{-p,m_z}^{\mathbf{bs},\lambda} \vert^2$. In black (right axis), plots of the backscattered cross section. The left and right images have been done for $ \mathbf{E}_{1,7}^{\lambda} $ and $\mathbf{E}_{1,8}^{\lambda}$ incident beams respectively. All simulations are done in a Mie theory regime, \textit{i.e.} the substrate has not been taken into account.  \label{fig:dual2}}
\end{figure}

Here, it is again observed that the AM plays a key role. In Figure \ref{fig:dual2} we show the same plot as in Figure \ref{fig:dual} when the incoming beams have $m_z=7,8$ and $p=1$. We see that for these two cases, it is no longer true that the minima of the backscattering are related to approximate maxima of helicity conservation. Still, as the value of the AM carried by the incoming beam increases, the helicity is better conserved - an effect which was already predicted in \cite{Zambrana2013OE}. This can be seen if we compare the red (left) axis of Figures \ref{fig:dual} and \ref{fig:dual2}, where it is seen that the conservation of helicity is increased by two orders of magnitude. 

In conclusion, we have experimentally shown that the AM and the helicity of light play a crucial role in the light scattering of a spherical particle. We have observed that by increasing the AM of the illumination, not only the scattering response can be tuned and sharpened, but also the level of conservation of helicity by a sphere can be improved by orders of magnitude. To show that, the backscattering of a single TiO$_2$ spherical particle deposited on a glass substrate has been experimentally measured. All experimental findings have been supported by theoretical predictions. The presented findings provide deeper insight of the role of helicity and angular momentum of light into the linear interaction between light and individual Mie scatterers. Moreover, our results depict an alternative way to potentially address single multipolar resonances for systems whose sizes are of the order of the wavelength or smaller. 

\section{Methods}
\small
\textbf{Experimental details}\\
\underline{Tightly focused beam engineering}: Tuning the wavelength of the laser introduces some experimental challenges. In order to maintain a perfect circular polarization over the whole spectral range of the experiment, the QWP needs to change its orientation at every single wavelength. This is controlled with a motorized system. Also, since the SLM position is static but its phase depends on the wavelength, a small correction has to be given at every single wavelength so that the optical singularity is well-aligned with the propagation axis of the set-up. Furthermore, due to the chromatic aberrations of the MO, the focal plane of the MO needs to be re-assessed at each wavelength. A motorized translation stage is used for this matter. Last but not least, in order to maintain the cylindrical symmetry of the system, the particle needs to be re-centered with respect to the incident beam for every change in wavelength or AM. This is achieved by looking at the CCD camera in real time and correcting the position of the particle with the nanopositioning system.\\
\underline{Backscattering measurements}: In order to measure the backscattering, three different power measurements have been carried out: $P_0,P_p,P_s$. $P_p$ is the power signal of the backscattering of the particle. $P_s$ is the signal of the backscattering of the substrate when the particle is displaced $10\mu m$ from it. And $P_0$ is the noise measured by the CCD camera when the laser is switched off but all the rest of devices are on. The three measurements are done using a CCD camera and integrating over all the relevant pixels. Great care has been taken not to saturate the CCD camera. In order to do that, each wavelength measurement is associated to an exposure time ($E$) and a neutral density filter value ($F$). Then, if a signal $P_i$ (with $i=0,p,s$) is obtained after integrating the values of all the active pixels of the CCD, the real signal is evaluated as:
\begin{equation}
P_{i'}=P_i/(E\cdot F)
\label{eq:signal}
\end{equation}
The previous relation assumes a linear response of the system, which is the case in our set-up due to the moderate power of the laser $<50mW$. Taking all this into account, we compute:
\begin{equation}
I=\dfrac{P_{p'}-P_{0'}}{P_{s'}-P_{0'}}
\end{equation} 
which is a dimensionless number that characterizes the backscattering of the particle that is being probed, where each of the individual $P_0,P_p,P_s$ signal measurements has taken into account eq.(\ref{eq:signal})  \\

\textbf{Numerical simulations}\\
\begin{figure}[tbp]
\centering
\includegraphics[width=10cm]{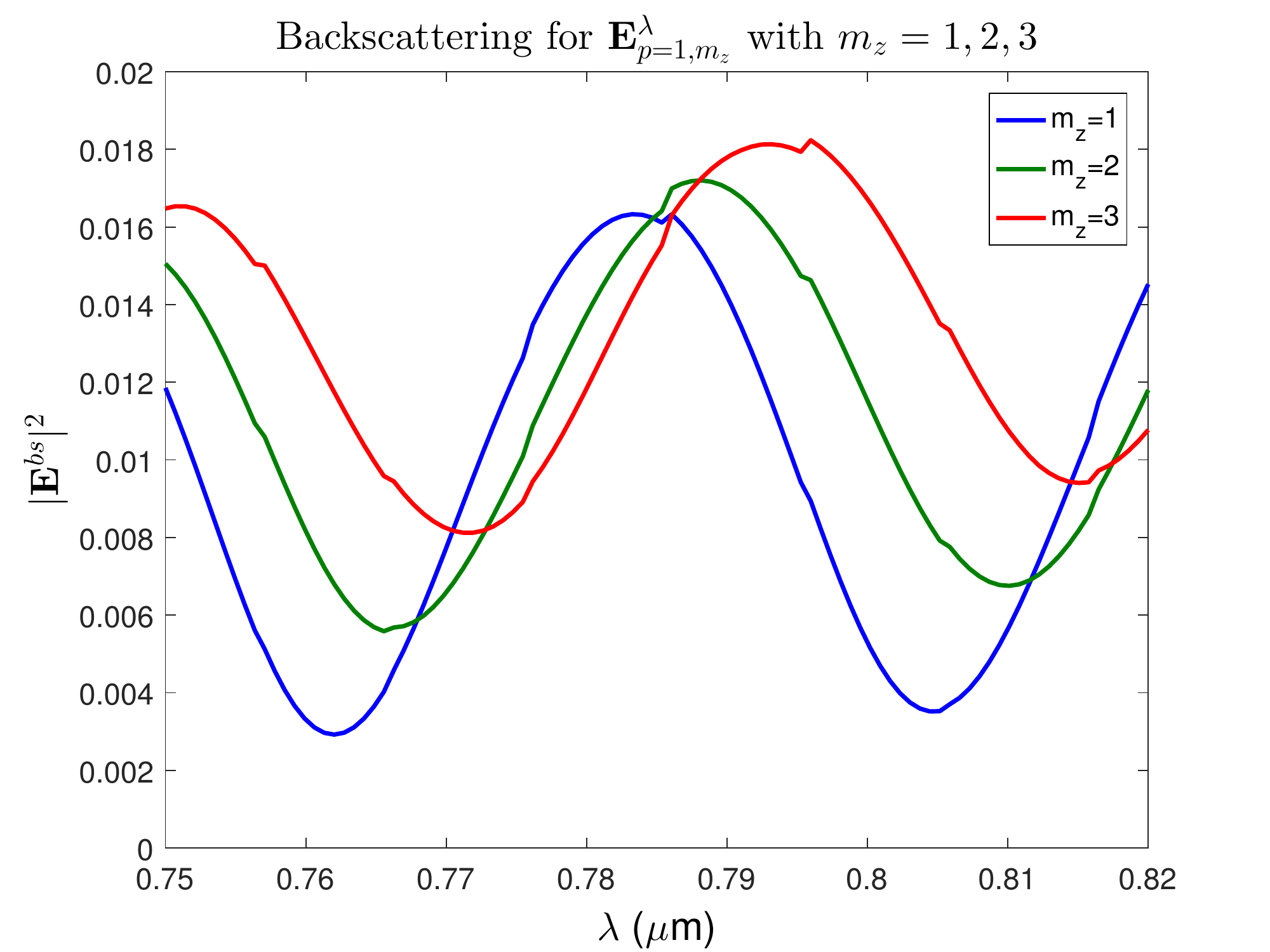} 
\caption{Theoretical backscattering with $l=0,1,2$.  \label{fig:subtle}}
\end{figure}
\underline{Backscattering simulations}: The backscattering simulations are done using an analytic Mie theory code. A sphere of $R=2 \mu m$ and $n_r=1.8$ embedded in air is used for all the simulations. The refractive index of the TiO$_2$ particle is 1.8 because they were in their amorphous phase \footnote{Private communication with \textit{MKNano}.}. The incident beam at the focal point $\Emp$ is decomposed into multipolar modes following the strategy in Ref. \cite{Zambrana2012}. All the $\Emp$ used in the simulations have the same beam waist before focusing, as well as a unitary intensity on a transverse plane. Given this set of incident beams, we obtain the scattered field. We integrate the scattered field on a transverse plane $1\mu m$ behind the particle. The plane has a surface of $10 \times 10 \mu m^2$. The integration of the fields in this transverse plane gives us the backscattering of the particle for a certain $\lambda$. The result is repeated over different $\lambda$s.\\
\underline{Shift of backscattering oscillations}: Here we show the results of some numerical simulations using a single TiO$_2$ sphere embedded in air with $R=2\mu m$ and $n_r=1.8$. The incoming beams are $\Emp$ for $\lambda=750-810$nm, $p=1$, and $m_z=1,2,3$. The simulations are done following the procedure described above. It is clearly observed that the backscattering maxima and minima shift. While the minimum for the beam with $m_z=1$ ($l=0$) is at $\lambda_m \approx 760$nm, the minima of the beams of $m_z=2,3$ appear to be shifted to 765 and 772 nm respectively. This feature highlights the fact that the spectral measurements shown in Figures \ref{fig:scan_l} and \ref{fig:scan_r} are oscillations, and not Mie resonances. Note that Mie resonances cannot be spectrally tailored with an incoming beam, as its value is fixed by the geometry and the material properties of the particle \cite{Zambrana2015}. That is, if the minimum that appears in Figure \ref{fig:subtle} was a resonances, it would not shift due to a change in the incoming beam.

\section{Acknowledgments}
This work was funded by the Australian Research Council Discovery Project DP160103332.


%

\end{document}